\documentclass{article}
\def\RR{{$RR$}}
\def\calN={{\cal N}}
\def\nn{\nonumber}

\def\ttt{{\tilde t}}

\def\tq{{\tilde q}}
\def\hm{{\widehat m}}

\def\ie{{\it i.e.}}
\def\half{{{1 \over 2}}}

\def\pmb#1{\setbox0=\hbox{#1}%
 \kern-.025em\copy0\kern-\wd0
 \kern.05em\copy0\kern-\wd0
 \kern-.025em\raise.0433em\box0 }
\font\cmss=cmss10
\font\cmsss=cmss10 at 7pt
\def\rlx{\relax\leavevmode}
\def\Cop{\relax\,\hbox{$\kern-.3em{\rm C}$}}
\def\Rop{\relax{\rm I\kern-.18em R}}
\def\Nop{\relax{\rm I\kern-.18em N}}
\def\Pop{\relax{\rm I\kern-.18em P}}

\def\Zop{\rlx\leavevmode\ifmmode\mathchoice{\hbox{\cmss Z\kern-.4em Z}}
 {\hbox{\cmss Z\kern-.4em Z}}{\lower.9pt\hbox{\cmsss Z\kern-.36em Z}}
 {\lower1.2pt\hbox{\cmsss Z\kern-.36em Z}}\else{\cmss Z\kern-.4em
 Z}\fi}

\def\m{{m}}

\def\bbbone {{\mathchoice {\rm 1\mskip-4mu l} {\rm 1\mskip-4mu l}
{\rm 1\mskip-4.5mu l} {\rm 1\mskip-5mu l}}}

\usepackage{fortschritte}
\def\beq{\begin{equation}}                     %
\def\eeq{\end{equation}}                       %
\def\bea{\begin{eqnarray}}                     
\def\eea{\end{eqnarray}}                       
\begin {document}                 
\def\titleline{
%
%
%
D-branes in a plane-wave background           
}
\def\email_speaker{
{\tt 
%
%
mrg@mth.kcl.ac.uk                            
}}
\def\authors{
%
%
%
%
%
M.R. Gaberdiel\1ad\sp, M.B. Green\2ad        
}
\def\addresses{
%
%
%
%
\1ad 
Department of Mathematics \\                %
King's College, London     \\               %
Strand, London, WC2R 2LS, UK \\             %
\2ad                                        %
Department of Applied Mathematics and Theoretical Physics \\
Cambridge University \\                     
Wilberforce Road, Cambridge, CB3 0WA, UK    
                                            %
                                            %
}
\def\abstracttext{
%
%
The D-branes of the maximally               
supersymmetric plane-wave                   
background are described.                   
}
\large
\makefront
\section{Introduction}

The study of type IIB superstring theory in plane wave backgrounds
is a fertile area for investigating the properties of string theory with
nontrivial Ramond--Ramond (\RR) condensates. One of the best studied
examples is the Penrose limit of $AdS_5\times S^5$ \cite{hulletal},
for which the dual CFT has a corresponding limit \cite{maldetal}.  
In the original $AdS_5\times S^5$ background this correspondence is
difficult to check quantitatively since string theory in the 
$AdS_5\times S^5$ background has not, so far, proved tractable. On the
other hand, the plane-wave theory can be formulated as a free
two-dimensional field theory, at least in the light-cone gauge, and is
therefore exactly solvable \cite{met,mt}.   

In this talk we shall describe the construction of supersymmetric
D-branes for this string theory background. Since the world-sheet
theory in light cone gauge is not conformally invariant, a number of
modifications arise relative to the usual conformal field theory
discussion. In particular, the functions that enter in the
calculation of the cylinder amplitudes are not standard $\theta$
functions, and the open-closed consistency condition follows from
rather non-trivial modular transformation properties of deformed
$\theta$ functions.  

This talk is based on \cite{bgg} and \cite{gg1}.

\section{Notation and review}

The $pp$-wave background is a ten-dimensional space-time with metric,
\beq
ds^2 = 2 dx^+ dx^- - \mu^2 x^I x^I dx^+ dx^+ + dx^I dx^I \, ,
\label{metpp}
\eeq
where $x^{\pm} =  (x^0 \pm x^9)/\sqrt{2}$ and $I=1,\dots,8$.
The five-form \RR\ field strength is given by
\beq
F_{+ 1234} = F_{+ 5678} = 2\, \mu\, ,
\label{fback}
\eeq
where $\mu$ is a constant. In light-cone gauge where 
$x^+ = 2 \pi\,  p^+ \tau$, the Lagrangian describes eight massive free
scalar and eight massive free fermion fields
\beq
{\cal L} = {1\over 4\pi} \left( \partial_+ x^I \partial_- x^I
- m^2 (x^I)^2  \right)
+ {i \over 2\pi}\left(S^a\partial_+ S^a + \tilde S^a \partial_- \tilde S^a
- 2m\, S^a\, \Pi_{ab} \, \tilde S^b \right)\, ,
\label{lcact}
\eeq
where $S^a$ and $\tilde S^a$ are $SO(8)$ spinors of the same chirality
and $\Pi = \gamma^1 \gamma^2\gamma^3 \gamma^4$. The mass parameter
$m$ is defined by $m=2 \pi p^+ \mu$. The $8\times 8$ matrices,
$\gamma^I_{a\dot b}$ and $\gamma^I_{\dot ab}$, are the
off-diagonal  blocks of the $16\times 16$ $SO(8)$ $\gamma$-matrices
and couple $SO(8)$ spinors of opposite chirality.  The presence of
$\Pi$ in the fermionic sector of the lagrangian breaks the symmetry
from $SO(8)$ to $SO(4)\times SO(4)$.

The modes of the transverse bosonic
coordinates, $x^I$, of a string in type IIB $pp$-wave light-cone gauge
string theory  \cite{met} are $\alpha^I_k$ and $\tilde\alpha^I_k$, where
$I$ is a vector index of $SO(8)$, and $k\in\Zop$ with $k\ne 0$. These
modes satisfy the commutation relations
\beq
[\alpha^I_k,\alpha^J_l]  = \omega_k \, \delta^{IJ}\, \delta_{k,-l}
\,, \qquad
[\alpha^I_k,\tilde\alpha^J_l]  = 0 \, , \qquad
[\tilde\alpha^I_k,\tilde\alpha^J_l] = \omega_k\, \delta^{IJ}\,
\delta_{k,-l} \,,
\label{boscom}
\eeq
where
\beq
\omega_k = {\rm sign}(k)\, \sqrt{k^2 + \m^2}\qquad |k|>0\, .
\label{omegadef}
\eeq
In addition there are the bosonic zero modes that describe the
centre of mass position $x^I_0$ and some generalised momentum $p^I_0$
with $[p_0^I,x_0^J] = - i \delta^{IJ}$. 
It is convenient to introduce the creation and annihilation operators
\beq
a^I_0 = {1\over \sqrt{2\m}} \bigl(p_0^I + i \m x_0^I\bigr) \,, \qquad
\bar{a}^I_0 = {1\over \sqrt{2\m}} \bigl(p_0^I - i \m x_0^I\bigr) \,,
\label{creani}
\eeq
in terms of which the commutation relation is then
$[\bar{a}_0^I,a_0^J] = \delta^{IJ}$.
The modes of the fermionic fields $S^a_k$ and $\tilde{S}^a_k$, where
$a$ is a spinor index of $SO(8)$ and $k\in\Zop$, satisfy the
anti-commutation relations 
\beq
\{ S^a_k,S^b_l\} = \delta^{ab}\delta_{k,-l} \,,  \qquad
\{ S^a_k,\tilde{S}^b_l\}  = 0 \,, \qquad
\{\tilde{S}^a_k,\tilde{S}^b_l\}  = \delta^{ab}\delta_{k,-l} \,.
\label{fermcom}
\eeq
It is convenient to introduce the zero-mode combinations
$\theta^a_0 = {1\over \sqrt{2}}(S^a_0 + i \tilde{S}^a_0)$ as well as  
$\bar\theta^a_0 = {1\over \sqrt{2}}(S^a_0 - i \tilde{S}^a_0)$, 
and further
\bea
\theta_R & =& \half (1 + \Pi) \theta_0 \,,  \qquad
\bar\theta_R = \half (1 + \Pi) \bar\theta_0 \,, \nn\\
\theta_L & =& \half (1 - \Pi) \theta_0 \,, \qquad
\bar\theta_L = \half (1 - \Pi) \bar\theta_0 \,.
\label{thetarl}
\eea
In light-cone gauge the thirty-two components of the supersymmetries
decompose into `dynamical' and `kinematical' components. The
`dynamical' supercharges, $Q_{\dot a}$  and $\tilde Q_{\dot a}$ 
($\dot a =1,\dots,8$), anticommute to give the light-cone hamiltonian 
\bea
2 \, p^+ H &=& \m \left(a^I_0\, \bar{a}^I_0
+ i\, S^a_0 \,\Pi_{ab}\, \tilde{S}^b_0 +4\right)
+ \sum_{k=1}^{\infty} \left[ \alpha^I_{-k}\alpha^I_k +
\tilde\alpha^I_{-k} \tilde\alpha^I_k
+ \omega_k \left(S^a_{-k} S^a_{k} + \tilde{S}^a_{-k} \tilde{S}^a_{k}
\right) \right]\nn\\
 &=& \m \left(a^I_0\, \bar{a}^I_0 + \theta_L^a \,\bar\theta_L^a +
\bar\theta_R^a \,\theta_R^a \right)
+  \sum_{k=1}^{\infty} \left[
\alpha^I_{-k}\alpha^I_k +
\tilde\alpha^I_{-k} \tilde\alpha^I_k
+ \omega_k \left(S^a_{-k} S^a_{k} + \tilde{S}^a_{-k} \tilde{S}^a_{k}
\right) \right]
\,.\nn\\
\label{lcham}
\eea
In addition the theory has sixteen `kinematical' supercharges  
$Q_a\equiv S^a_0$ and $\tilde Q_a\equiv \tilde{S}^a_0$, where 
$a=1,\ldots,8$. The dynamical supercharges commute with the light-cone
hamiltonian, but the kinematical supercharges do not.

\section{The construction of the boundary states}

D-branes in string theory can always be described by boundary
states. This description makes the coupling of the D-brane to
the closed string states of the theory manifest. 
In the following we shall only be discussing instantonic D-branes,
which are defined by the embeddings of euclidean  $(p+1)$-dimensional
world-volumes.\footnote{As is explained in \cite{bgg,gg1}, the
following discussion can be easily generalised for time-like branes.}
These are the cases in which the light-cone directions 
$x^\pm$ are orthogonal to the brane. We will adopt a notation,
following \cite{st}, in which these instantonic branes are denoted
$(r,s)$-branes ($r+s = p+1$), where $r$  and $s$ are the numbers of
Neumann directions associated with the two $SO(4)$ factors in the
transverse space. 

The boundary states are (up to important normalisations) uniquely
determined by the gluing conditions that describe how left- and
right-moving fields are related at the boundary. Here we shall only
consider D-branes that preserve half of the dynamical
supersymmetries, \ie\
\beq
\left( Q_{\dot a} + i \eta\, M_{\dot a \dot b}\,
{\tilde Q}_{\dot b}\ \right)
|\!| (r,s),\eta\,\rangle\!\rangle = 0\,,
\label{dynamicalnn}
\eeq
where the value of $\eta =\pm 1$ distinguishes a brane from an
anti-brane. The matrix $M$ is  given by
\beq
M_{\dot{a}\dot{b}} =
\left( \prod_{I\in{\cal N}} \gamma^I \right)_{\dot{a}\dot{b}}\,,
\label{Mdot}
\eeq
where the product extends over the Neumann directions ${\cal N}$. In
order for the boundary state to define a standard D-brane one requires
in addition that the bosonic degrees of freedom satisfy 
\bea
\left( \alpha^I_k - \tilde\alpha^I_{-k} \right)
|\!| (r,s), {\bf y}_t\,\rangle\!\rangle  & = & 0\,,
\qquad k\in\Zop\,, \nn\\
\left( \bar{a}^I_0 - a^I_0 + i \sqrt{2m} y_t^I \right)
|\!| (r,s), {\bf y}_t\,\rangle\!\rangle &\equiv &
- i\sqrt{2m}\, (x_{0}^I - y_t^I) \,
|\!| (r,s), {\bf y}_t\,\rangle\!\rangle = 0 \,,
\label{boundarycond}
\eea
for each Dirichlet direction $I$, as well as 
\bea
\left( \alpha^J_k + \tilde\alpha^J_{-k} \right)
|\!| (r,s), {\bf y}_t\,\rangle\!\rangle  & = & 0\,,
\qquad k\in\Zop\,, \nn\\
\left( \bar{a}^J_0 + a^J_0 \right)
|\!| (r,s), {\bf y}_t\,\rangle\!\rangle & = & 0
\label{boundarycondN}
\eea
for each Neumann direction $J$. (For simplicity we are considering
here the case without Wilson lines.) It is easy to see that
(\ref{dynamicalnn}), together with (\ref{boundarycond}) and 
(\ref{boundarycondN}) implies that 
\beq
\left( S^a_0 + i\, \eta\, M_{ab}\,\tilde{S}^b_0 \right)
|\!| (r,s),{\bf y}_t\,\rangle\!\rangle = 0\,,
\label{kinematical}
\eeq
where $M_{ab}$ is the same product of $\gamma$ matrices as in
(\ref{Mdot}). This last equation implies that a complex combination of
the kinematical supersymmetries is preserved by the boundary state. 
The structure of the boundary states depends now crucially on the
choice of $M$.

\noindent {\bf Class I}: The first class is the one that was studied 
in \cite{bp,dp,bgg} and arises when the matrix $M_{ab}$ satisfies
\beq
\Pi M \Pi M = -1\, ,
\label{classone}
\eeq
where $\Pi =\gamma^1\gamma^2\gamma^3\gamma^4$. From a supergravity
point of view, these branes were analysed in \cite{st}. The condition 
(\ref{classone}) is equivalent to $|r-s|=2$. The branes of this kind
satisfy the standard fermionic gluing condition 
\beq
\left( S^a_n + i\, \eta\, M_{ab}\,\tilde{S}^b_{-n} \right)
|\!| (r,s),{\bf y}_t\,\rangle\!\rangle = 0\,.
\label{kinematicaln}
\eeq
They satisfy (\ref{dynamicalnn}) if $y_t^I=0$ for the Dirichlet
directions, and the corresponding open string preserves eight
components (\ie\ half) of both the dynamical and kinematical
supersymmetries.  A characteristic feature of this class is that the
kinematical conditions (\ref{kinematical}) are not preserved as a
function of $x^+$ since the commutator with the light-cone hamiltonian
has the form 
\beq
[H,Q_a + i \eta\, M_{ab}\,{\tilde Q}_b] =
{m\, \eta \over 2 p^+} (\Pi M^t)_{ab}
\left(Q_b  - i \eta\, M_{bc}\,{\tilde Q}_c \right) \,.
\eeq
In this case the open-string theory has a mass term in its hamiltonian
of the form $S_0 M\Pi S_0$ \cite{dp}, and the ground state is an
unmatched boson. In particular, the open string one-loop amplitude
(and the corresponding closed string cylinder diagram) therefore does
not vanish. 

\noindent {\bf Class II}: The second class arises when  the matrix
$M_{ab}$ satisfies
\beq
\Pi M \Pi M = 1\, ,
\label{classtwo}
\eeq
a possibility that was not considered in \cite{bp,dp,bgg} but
arose in the supergravity analyses of \cite{st,bmz} and was later
analysed in detail in \cite{gg1} (the open string description was
independently worked out in \cite{st2}). In this
case $|r-s|=0,4$. The only branes of this type that satisfy
(\ref{dynamicalnn}) are the $(0,0)$ brane at an arbitrary
position\footnote{The same also holds for the $(4,4)$ brane, but the
corresponding boundary state does not have a sensible flat space limit
and is therefore probably inconsistent.}, for which the fermionic
gluing conditions are   
\beq\label{fermions1}
\left(S^a_n + i \eta\, R_n^{ab}\,
 \tilde{S}^b_{-n} \right)
|\!| (0,0),{\bf y}\,\rangle\!\rangle = 0  \,,
\eeq
and $R_n$ is the matrix
\beq\label{rdef}
R_n = {1\over n} \left( \omega_n \bbbone - \eta\, m \Pi
\right) \,.
\eeq
In addition, the $(4,0)$ and $(0,4)$ branes satisfy (\ref{dynamicalnn}) 
provided that the Neumann boundary conditions (\ref{boundarycondN})
are suitably modified (see \cite{gg1}). The corresponding open strings
preserve half of the dynamical supersymmetries, but none of the
kinematical supersymmetries. However, in contradistinction to class I,  
\beq
[H,Q_a + i \eta\, M_{ab}\,{\tilde Q}_b]
= - {m \, \eta \over 2 p^+} (\Pi M^t)_{ab}
\left( Q_b + i \eta\, M_{bc}\, {\tilde Q}_c \right)\,,
\eeq
and thus the kinematical conditions (\ref{kinematical}) {\it are}
preserved as a function of $x^+$. In this case
$S_0 M\Pi S_0\equiv 0$ and the open-string mass term vanishes, thus
giving true fermionic zero modes (see also \cite{bpz}). The ground
states then form a degenerate supermultiplet, and  the one-loop open
string amplitude (as well as the corresponding closed string cylinder
diagram) vanishes.

\section{Open-closed duality}

One of the most important consistency conditions for D-branes is the
so-called open-closed duality relation. It requires that the cylinder
diagram, that describes the interaction energy between two D-branes,
can be evaluated in two different ways. From the closed string point
of view, the cylinder diagram is the tree-level diagram describing the
exchange of closed string states between two external boundary
states. On the other hand, the diagram can also be interpreted as a
one-loop open string diagram. These two points of view are related by
exchanging the roles of the world-sheet parameters $\tau$ and
$\sigma$. This transformation replaces the modular parameter $t$ of
the cylinder by $\ttt=1/t$; it also replaces the mass-parameter $m$ by
$\hm=mt$. (See \cite{bgg,gg1} for more details.)

In the usual flat space background, the open-closed duality condition
is satisfied since the cylinder amplitude can be written in terms of
standard $\theta$ functions that have well understood transformation
properties under the modular transformation $t\mapsto\ttt$. For the
case of the plane-wave background, on the other hand, the cylinder
amplitudes involve non-trivial deformations of $\theta$ functions. In 
the case of the class I branes\footnote{The functions that arise for
the class II branes are yet more complicated; they are described in
detail in \cite{gg1}, where their modular properties are also 
discussed.}, the relevant functions are  
\bea
f_1^{(\m)}(q) & =& q^{-\Delta_\m} (1-q^\m)^{{1\over 2}}
\prod_{n=1}^{\infty} \left(1 - q^{\sqrt{\m^2+n^2}}\right) \,,
\label{fdef}\\
f_2^{(\m)}(q) & = &q^{- \Delta_\m} (1+q^\m)^{{1\over 2}}
\prod_{n=1}^{\infty} \left(1 + q^{\sqrt{\m^2+n^2}}\right) \,,
\label{f2def}\\
f_3^{(\m)}(q) & =& q^{-\Delta^\prime_\m}
\prod_{n=1}^{\infty} \left(1 + q^{\sqrt{\m^2+(n-1/2)^2}}\right) \,,
\label{f3def}\\
f_4^{(\m)}(q) & =& q^{-\Delta^\prime_\m}
\prod_{n=1}^{\infty} \left(1 - q^{\sqrt{\m^2+(n-1/2)^2}}\right) \,,
\label{f4def}
\eea
where $q=e^{-2\pi t}$, and $\Delta_\m$ and $\Delta^\prime_\m$ are
given as  
\bea
\Delta_\m & =& -{1\over (2\pi)^2} \sum_{p=1}^{\infty}
\int_0^\infty ds \, e^{-p^2 s} e^{-\pi^2 \m^2 / s} \,, \cr
\Delta^\prime_\m & =& -{1\over (2\pi)^2} \sum_{p=1}^{\infty}
(-1)^p \int_0^\infty ds \, e^{-p^2 s} e^{-\pi^2 \m^2 / s}\,.
\label{Deltadef}
\eea
The quantities $\Delta_m$ and $\Delta^\prime_m$ are the Casimir
energies of a single (two-dimensional) boson of mass $\m$ on a
cylindrical world-sheet with periodic and anti-periodic boundary
conditions, respectively. For $\m=0$, $\Delta_\m$ and
$\Delta^\prime_\m$ simplify to the usual flat-space values,
\bea
\Delta_0 & =& -{1\over (2\pi)^2} \sum_{p=1}^{\infty} {1\over p^{2}} =
- {1\over 24} \,, \cr
\Delta^\prime_0 & =& -{1\over (2\pi)^2}
\sum_{p=1}^{\infty} {(-1)^p \over p^{2}} =  {1\over 48}\,.
\label{mzero}
\eea
Thus $f_2^{(\m)}(q)$, $f_3^{(\m)}(q)$ and $f_4^{(\m)}(q)$ simply
reduce to the standard $f_2(q)$, $f_3(q)$ and $f_4(q)$ functions 
\cite{polcai} as $\m\rightarrow 0$.

For the case of the class I branes the open-closed consistency
condition is then a consequence of the remarkable transformation
properties of these functions \cite{bgg}
\beq
f_1^{(\m)}(q) = f_1^{(\hm)}(\tq)\,, \qquad f_2^{(\m)}(q) =
f_4^{(\hm)}(\tq)\,, \qquad f_3^{(\m)}(q) = f_3^{(\hm)}(\tq)\,,
\label{beautiful}
\eeq
where $\tq=e^{-2\pi \ttt}=e^{-2\pi/t}$ and $\hm=mt$. 
In the limit $\m\rightarrow 0$ the second and third equations in
(\ref{beautiful}) reproduce standard $\theta$-function identities.
The identity for $f_1$ (or $\eta$) can also be derived from the first
equation of (\ref{beautiful}). In fact, both sides of
the first equation tend to zero as $\m\rightarrow 0$ since
$(1-q^\m)^{1\over 2} =\sqrt{2\pi t \m} + {\cal O}(\m)$ and
$(1-\tilde{q}^{\,\widehat{\m}})^{1\over 2}=\sqrt{2\pi \m}+{\cal O}(\m)$. 
Thus, after dividing the first equation by $\sqrt{\m}$, the limit
$\m\rightarrow 0$ becomes
\beq
f_1(q) = t^{-\half} f_1(\tilde{q}) \,,
\label{etatrans}
\eeq
thus reproducing the standard modular transformation property of the
$f_1$ (or $\eta$) function.

{\bf Acknowledgement} MRG would like to thank the organisers of the
symposium for organising a very pleasant conference, and for giving him
the opportunity to speak. He is grateful to the Royal Society for a
University Research Fellowship.  We also
acknowledge  partial support from the PPARC Special Programme Grant
`String Theory and Realistic Field Theory', PPA/G/S/1998/0061 and the
EEC contract HPRN-2000-00122.


\end{document}